\documentclass[prb,twocolumn,showpacs,floatfix,amsmath,amssymb,superscriptaddress]{revtex4}
\usepackage[pdftex]{graphicx}
\usepackage{latexsym}
\usepackage{graphicx}
\usepackage{times}
\usepackage{amsmath}
\usepackage{dcolumn}
\usepackage{subfigure}
\usepackage{latexsym,amsmath,amssymb,bm,euscript}
\usepackage{amsfonts}
\usepackage{hyperref}

\usepackage{color}

\newcommand{\be}{\begin{equation}}
\newcommand{\ee}{\end{equation}}

\newcommand{\eps}{\epsilon}

\newcommand{\BiSe}{Bi$_2$Se$_3$ }

\begin{document}

\title{A thermodynamic measure of the Magneto-electric coupling in the 3D topological insulator}  
\author{Doron L. Bergman}
\affiliation{Physics Department, California Institute of Technology, MC 114-36, 1200 E. California Blvd., Pasadena, CA 91125}
\author{Gil Refael}
\affiliation{Physics Department, California Institute of Technology, MC 114-36, 1200 E. California Blvd., Pasadena, CA 91125}
\affiliation{
Dahlem Center for Complex Quantum Systems and Fachbereich Physik,
Freie Universit\"at Berlin, 14195 Berlin, Germany
}
 
\date{\today} 
 
\begin{abstract}
We show that the magneto-electric coupling in 3D (strong) topological insulators is related to a second derivative
of the bulk magnetization. The formula we derive is the non-linear response analog of the Streda formula 
for Hall conductivity ( P. Streda, J. Phys. C: Solid State Physics, 15, 22 (1982) ),
which relates the Hall conductivity to the derivative of the magnetization with respect to chemical potential.
Our finding allows one to extract the magneto-electric coefficient by measuring the magnetization,
while varying the chemical potential and one more perturbing field, a unique method never attempted before in the experimental search for the magneto-electric effect.
The relation we find also makes transparent the effect of disorder on the 
magneto-electric response, which occurs only through the density of states,
and has no effect when the system is gapped.
\end{abstract} 
 
\maketitle 

The Quantum Hall effect\cite{IQHE1} (QHE) was the first experimental instance where 
a transport coefficient (the Hall conductivity in $d=2$) was quantized\cite{Hastings:2009}.
Finding an analog system in $d=3$ had remained
an unrealized dream of the condensed matter community,
for many years.
Recently, however, this has changed. The discovery of the topological insulator\cite{Kane:2005A,Kane:2005B,Bernevig:2006},
and specifically the 3D strong topological insulator (STI)\cite{Fu:2007,Moore:2007},
have finally realized the dream of a $d=3$ analog of the QHE.
In the context of the QHE, Streda et al.\cite{Streda:1977,Streda:1982} proved an 
insightful relation between the Hall conductivity, and a second derivative of a 
thermodynamic potential
\be\label{ideal_Streda}
\sigma_{\textrm{Hall}} = \frac{1}{2}\left( \sigma_{x y} - \sigma_{y x} \right) =
- \frac{\partial \rho}{\partial B} \Big|_{\mu}
= - \frac{\partial M}{\partial \mu}\Big|_{B}
\; ,
\ee
where $\rho$ is the charge density, 
$M$ the orbital magnetization per unit volume (perpendicular to the 2D system),
$B$ is the external magnetic field, and $\mu$ the chemical potential.
Here and throughout the manuscript we use units where $c=e=\hbar=1$.

Motivated by the analogy between the $d=2$ QHE and $d=3$ STI, 
in this paper we show that the magneto-electric coefficient characterizing the STI, is also related to a thermodynamic derivative. We use an alternative formulation of magneto-electric effect, as a current response defined by
\be\label{new_formulation}
{\bf J} = \frac{\chi}{2\pi} \nabla \phi  \times {\bf E}
\; ,
\ee
where $\phi$ is an external field odd under inversion and time-reversal, and
$\chi$ is the magneto-electric response coefficient, and show that
\be\label{grand_result}
\chi = 
-\frac{2 \pi}{3}
\frac{\partial^2 \rho}{\partial B^a \partial (\partial_a \phi)} 
= 
-\frac{2 \pi}{3}
\frac{\partial^2 M_a}{\partial \mu \partial (\partial_a \phi)}
\; .
\ee 
There are a number of merits to this result, similar to those of \eqref{ideal_Streda}.
First, \eqref{Streda} suggests we can measure the Hall conductivity in the QHE, by doing a thermodynamic measurement - vary the chemical potential through a back gate, and measure the magnetization of the sample. Calculating the derivative of the magnetization with respect to the gate voltage should give $\sigma_{\textrm{Hall}}$. Similarly, with \eqref{grand_result} we can measure $\chi$ by simultaneously varying the chemical potential $\mu$ and the gradient $(\partial_a \phi)$, while measuring the magnetization of the sample.
Second, the effects of disorder on \eqref{ideal_Streda} and \eqref{grand_result} are entirely included in the density of states (DOS) $D(\eps)$ through the particle density $\rho = \int d\eps f(\eps) D(\eps)$, where $f(\eps)$ is the Fermi Dirac distribution. Disorder will reduce the effective gap in the spectrum, compared with the clean limit\cite{Shindou:prb2009}, but otherwise will not change anything for an insulator, as long as the chemical potential remains in the gap.

The magneto-electric effect\cite{Qi:prb2008,Essin:2009} is usually formulated
as an anomalous term appearing in the action for the electro-magnetic fields
in an insulator
$S_{E B} =  
\frac{1}{2 \pi} \int d{\vec r} dt P_3 {\vec E} \cdot {\vec B} $,
with $P_3$ the magneto-electric coefficient. 
Under inversion, ${\vec E}$ is odd, and under time reversal ${\vec B}$ is odd,
so $P_3$ is odd under both. Ref.~\onlinecite{Qi:prb2008} showed that $P_3$ 
takes on values modulo $1$, and so in a material with either time reversal or inversion 
symmetry (or both), it can take on the values $P_3 = 0,\frac{1}{2}$. The value
$P_3 = \frac{1}{2}$ then characterizes the STI.
This value can in principal be measured\cite{Essin:2009} as
\be\label{P3_formula}
P_3 \delta_{a b} = 
2\pi \frac{\partial M_a}{ \partial E^b} = 
2\pi \frac{\partial P_a}{ \partial B^b}
\; ,
\ee
where the Latin letters $a,b = x,y,z$ denote spatial directions.
Here and throughout the manuscript we will use the Einstein summation convention.

At this point in time, a number of materials have been identified as 
topological insulators using spectroscopy to characterize their unique surface states\cite{
Hsieh:2008,Xia:2009,Hsieh:prl2009,Hsieh:2009,Hsieh:2010}
(an odd number of Dirac points).
However, thus far measuring the magneto electric coefficient has proved challenging.
The materials, by and large, have proven rather poor insulators, with significant 
carrier concentration, in some cases even a bulk Fermi surface appearing\cite{Analytis:2010,Eto:2010,Paglione:2012,Paglione:natphys2012,Paglione:2011,Steinberg:prb2011,Steinberg:2010}.
In a bulk metal DC electric fields are screened, making it impossible to measure
\eqref{P3_formula} directly.

Many other indirect ways have been proposed to detect the magneto-electric effect, most of them relying on the surface states in the STI \cite{Qi:RMP,MacDonald:prl2010,Lan:prb2011,Karch:prl2009,LaForge:prb2010,Maciejko:prl2010,Hankiewicz:prb2011,Garate:prl2010,Drew:prb2010,Essin:2009,Nagaosa:prb2010,Jenkins:prb2010,Qi:prb2008}, and on the Witten effect\cite{Witten:1979,Franz:prb2010,Qi:prb2008}.
The magneto-electric effect at the surface appears as a consequence of the material boundary,
where $P_3$ has a sharp jump, and the Witten effect appears as a consequence of a flux monopole entering the STI.
Our motivation in this work is to generalize the measurable effects of the 
magneto-electric coupling, beyond the surface and Witten effects, to response in the bulk.
This can be achieved, as we explain below, by using the field $\phi$ imitating the 4th vector potential component in 4D $A_4$, and replacing $P_3$.
With this extra field $\phi$, we reformulate the magneto-electric effect with \eqref{new_formulation}, and find the thermodynamic relation \eqref{grand_result} similar to 
\eqref{ideal_Streda}. Formulated in this way, we will also be able to talk about magneto-electric effects in gapless systems,
as was recently explored in Refs.~\onlinecite{Bergman:prl2011,Barkeshli:2011}.


\begin{figure}
	\centering
		\includegraphics[width=3.0in]{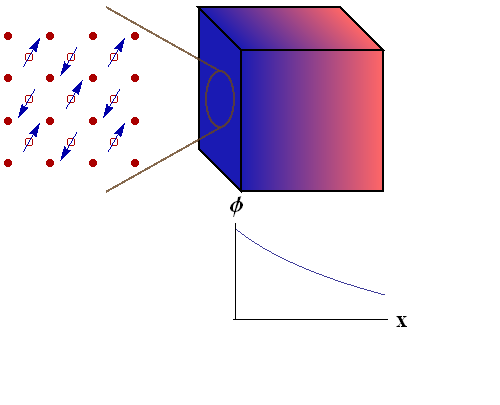}
	\caption{Measuring the topological magneto-electric response with anti-ferromagnetic 
	(AFM) order assuming the role of the auxiliary field $\phi$, odd under both inversion 
	and time-reversal. At the microscopic level, some of the ions are non-magnetic
	(denoted by full circles), while others are magnetic (denoted by open circles), 
	and form AFM order (denoted by the arrows). A slow gradient in $\phi$ in combination 
	with the electromagnetic field generate a magneto-electric response. The $\phi$ gradient
	indicated in the graph at the bottom is represented in the image of the material by 
	a change of shade (light red to light blue).}
	\label{fig:Streda_image}
\end{figure}

Since the STI is the $d=3$ analog of the QHE, we look to $d=2$ for inspiration.
First, the analog of magneto electric coupling in a gapless metal\cite{Bergman:prl2011,Barkeshli:2011} is the intrinsic anomalous Hall 
effect (AHE)\cite{Luttinger:1954,Jungwirth:prl2002,Onoda:2002,AHE_RMP}: in metals, where time-reversal is broken,
a large clean limit contribution to the Hall conductivity appears.
The general expression for the intrinsic AHE contribution to the Hall conductivity
becomes quantized once the chemical potential is in a gap of the band structure,
giving the integer QHE. In a metal the effect of disorder cannot be neglected,
and there are additional contributions to the Hall conductivity\cite{AHE_RMP}.
The interplay of the intrinsic AHE and disorder can be understood using
the so-called Bastin formula\cite{Bastin:1971,Streda:1977,Bruno:prb2001} for the 
electric conductivity tensor, which distinguishes between contributions
from states at the Fermi energy (the Fermi surface) $\sigma^{I}$ and 
contributions from all filled states (the Fermi sea) $\sigma^{II}$.
In an insulator, there are no bulk states at the Fermi energy, giving $\sigma^{I} = 0$.
Streda et al.~\onlinecite{Streda:1977,Streda:1982}
were able to show that the $\sigma^{II}$ contribution can be related 
to a derivative of the orbital magnetization
\be\label{Streda}
\sigma_{\textrm{Hall}}^{II} = 
- \frac{\partial \rho(\mu)}{\partial B} \Big|_{\mu}
= - \frac{\partial M}{\partial \mu}\Big|_{B}
\; .
\ee
Here $M,B$ are the orbital magnetization and magnetic field, respectively, ignoring Zeeman coupling to the electron spin. 
The formula holds in both gapped and gapless systems.
For an insulator, $\sigma^I = 0$, and the total Hall conductivity is reduced to \eqref{ideal_Streda}. 

The result \eqref{ideal_Streda} can be anticipated from the following considerations.
In an insulator, with no dissipative currents, the only currents possible are persistent currents
related to the orbital magnetization ${\bf J} = \nabla \times {\bf M}$.
The electric field is found from ${\bf E} = - \nabla \mu({\bf x})$.
Assuming the magnetization is an entirely \emph{local} function of the intensive thermodynamic quantities $M_a = M_a(T,{\bf B},\mu({\bf x}))$, we find
\be\label{Widom_formula}
J^a = \eps^{a b c} \partial_b M_c 
=
\eps^{a b c} \frac{\partial M_c}{\partial \mu} \frac{\partial \mu}{\partial x^b} 
= 
- \eps^{a b c} \frac{\partial M_c}{\partial \mu} E_b
\; ,
\ee
resulting in $\sigma_{xy} = - \frac{\partial M_z}{\partial \mu}$.
The second equality in \eqref{Streda} is due to a Maxwell relation\cite{Widom:1982,Streda:1983}.

As mentioned earlier, \eqref{Streda} indicates the $\sigma^{II}$ contribution can be found by doing a thermodynamic measurement.
However, the measured magnetization will include both the orbital and Zeeman contributions
to the magnetization, while \eqref{Streda} involves the \emph{orbital} 
magnetization alone.
For an insulator one can argue that the magnetization due to Zeeman coupling does not vary with chemical potential, and therefore measuring the derivative of the total magnetization, will give the 
same result as if we were measuring the orbital magnetization alone.
For a metal on the other hand, the Zeeman effect magnetization can depend on the chemical potential, for instance in Pauli paramagentism. Therefore, measuring the total magnetization will only give a quantitatively accurate measure of $\sigma^{II}$ in an insulator.
Still, it will be useful in finding qualitative differences - it will exhibit quantization 
when the bulk is gapped. 
While such a measurement is conceptually straightforward, in practice it is is more difficult
than measuring Hall effect through electric currents. However, it has been carried out\cite{Usher:2009,Wiegers:1998}. Next we will present the analog of \eqref{Streda} in the $d=3$ STI.

In order to deal with a possibly gapless spectrum, as well as with disorder, we will have to 
formulate the magneto-electric coupling in a slightly different way from Refs.~\onlinecite{Qi:prb2008,Essin:2009}. 
The magneto-electric effect relates the magnetization to an applied electric field
${\bf M} = \frac{P_3}{2\pi} {\bf E} $
equivalent to \eqref{P3_formula}. The Hall conductivity is well-defined for both an insulator and a metal
because it is defined through a current response to an external field
$J_x = \sigma_{x y} E_y$. For an insulator, ${\bf J} = \nabla \times {\bf M}$, 
and using Faraday's law $\nabla \times {\bf E} = - \partial_t {\bf B}$, we get
\be\label{form1}
{\bf J} = \frac{1}{2\pi} \nabla P_3  \times {\bf E} - \frac{P_3}{2\pi} \partial_t {\bf B}
\; .
\ee
The only DC (static) response comes from the first term
\be\label{form2}
{\bf J} = \frac{1}{2\pi} \nabla P_3  \times {\bf E}
\; .
\ee
Formulated in this way, we see $P_3$ should be regarded as an external field.

In fact, any inhomogeneous external field $\phi$ with the same symmetry 
properties of $P_3$, namely odd under inversion and time reversal, 
should suffice to generate such a current response. We therefore write
${\bf J} = \frac{\chi}{2\pi} \nabla \phi  \times {\bf E} $, which is just \eqref{new_formulation}. 
The response coefficient $\chi$ is the generalization of the magneto-electric coupling
in the insulator, and a more faithful analog of the Hall conductivity - it can now be 
defined and calculated for metals as well\cite{Bergman:prl2011}.
For an insulator, $\chi$ will be quantized, yet the precise value of $\chi$ depends 
on how the field $\phi$ is defined and couples to 
the system, and therefore by itself will \emph{not} attain a universal quantized value.
This is the one sacrifice we have to make in the alternate formulation of the magneto electric response.
On the other hand, it will prove a more robust quantity to measure, 
in a system that may be gapless, and most importantly it will
exhibit quantization in an insulator - the key qualitative feature we are after.

Next we will derive \eqref{grand_result}.
The derivation in the body of this manuscript is not rigorous,
and does not apply to gapless systems. It is presented here for the 
sake of brevity and clarity. In the supplementary material~\ref{supp_mat}
we will derive the result with some assumptions, while a general rigorous derivation 
is left for a future publication.
Using charge conservation $\partial_t \rho = - \nabla \cdot {\bf J}$,
as well as Faraday's law
$\nabla \times {\bf E} = - \partial_t {\bf B}$ we find from \eqref{new_formulation}
\be\label{form4}
\rho = - \frac{\chi}{2\pi} \nabla \phi \cdot {\bf B} 
\; .
\ee
The second term from \eqref{form1} has no contribution since $\nabla \cdot {\bf B}= 0$.
Taking $\partial_c \phi = h_c$, the differential form, \eqref{form4} becomes
\be\label{grand_result1}
\chi \delta_a^b = 
-2 \pi
\frac{\partial^2 \rho}{\partial B^a \partial h_b} 
= 
-2 \pi
\frac{\partial^2 M_a}{\partial \mu \partial h_b}
\; ,
\ee 
where we used the same Maxwell relation as in \eqref{Streda}.
Note that the magneto-electric response is found from the magnetization \emph{parallel}
to the direction of the auxiliary field gradient. This should not be surprising as the
magneto-electric effect should not care about whether the system is isotropic or not. 
In an isotropic system, the absence of any other directionality necessitates
this outcome. Contracting the $a,b$ indices in $d=3$, we arrive at \eqref{grand_result}.

A rigorous proof of \eqref{grand_result}, as well as a generalization to 
gapless systems can be derived with some effort.
Following standard response theory techniques\cite{Mahan}, we can
find the nonlinear response analog of the Bastin Formula
\be\label{nonlinear_Bastin}
\begin{split}
\chi^{I} = &
\frac{4 \pi}{4!} 
\eps^{a b c}
Re\left\{
Tr\left[
\delta(H)
\left( v_b G_R v_{\phi} - v_{\phi} G_R v_b \right) 
G_R v_c G_R v_a
\right]
\right\}
\\
\chi^{II} = & 2\pi
\frac{\eps^{\mu \nu \lambda \tau}}{4!}
\int_{-\infty}^{+\infty}
\frac{d\eps}{2\pi i} f(\eps)
\\ & 
Tr\left[ v_{\phi} G_R v_{\mu} G_R v_{\nu} 
G_R v_{\lambda} G_R v_{\tau} G_R \right]
+ c.c.
\; ,
\end{split}
\ee
where $v_a$ are the velocity operators,
$f(\eps)$ is the Fermi Dirac distribution,
$\delta(x)$ is the Dirac delta function,
and $G_R= \left[ \eps + i \delta - H \right]^{-1}$ 
is the retarded Green's functions, which can include any random potential.
The chemical potential is included in $H = \ldots - \mu$, so that the Fermi energy is 
at $\eps = 0$. The trace is over all degrees of freedom of the system - real 
(or momentum) space coordinates, and internal degrees of freedom.
Also, in the expression for the contribution $\chi^{II}$, all Green's functions depend on the frequency $\eps$, while in the expression for $\chi^{I}$, all Green's functions have
$\eps=0$.
Finally, the velocity $v_{\phi}$ is the conjugate operator to the auxiliary field
$H = H_0 + \int_{\bf x} \phi({\bf x}) v_{\phi}$.
The full details of this derivation we leave for a future publication.
In the supplementary material~\ref{supp_mat} we provide a limited derivation, appropriate 
for an insulator, with the field $\phi = h_a x^a$ coupled to a momentum independent $v_{\phi}$.

Much like in the case of conductivity, the form \eqref{nonlinear_Bastin} 
distinguishes between Fermi surface contributions $\chi^{I}$, which vanish for an insulator,
and the contribution $\chi^{II}$, which turns out to be the second derivative of the orbital
magnetization, satisfying \eqref{grand_result}
\be\label{grand_result2}
\chi^{II} = 
- \frac{2 \pi}{3}
\frac{\partial^2 M_a}{\partial \mu \partial h_a}
\; .
\ee 

As we already discussed in the introduction, our result \eqref{grand_result2},
much like the Streda formula \eqref{Streda}, indicates that $\chi$ can be obtained by measuring the 2nd derivative of the (orbital) magnetization.
As noted earlier, at least in the insulating case, the Zeeman contribution to the magnetization should not vary with the chemical potential, and measuring the 
full magnetization instead of the orbital magnetization alone, will yield the same result.
It is then conceptually straightforward to measure magnetization, and vary the chemical potential.
The auxiliary field gradient $h_a$, on the other hand, is at this point an abstract object
we defined for our theoretical needs. We turn our attention now to discussing how $\phi$
can be realized. First, given that $\phi$ must be odd under time reversal and inversion,
it can appear when anti-ferromagnetic (AFM) order is present in the material.
It is not unimaginable that a topological insulator material could be stuffed with 
magnetic atoms that realize AFM order in the material. Second, we need $\phi$ to vary (slowly) in space, as illustrated in Fig.~\ref{fig:Streda_image}.
This can occur naturally in AFM order, as it tends to form magnetic domains.
More difficult will be controlling and varying the strength of the AFM field.
This can be done by changing the temperature of the system, and for better control of it,
to be sufficiently close to the critical temperature of the AFM order.
We mention in passing that if the material lacks inversion symmetry, a field breaking time reversal alone should suffice to generate $\phi$. However, a vast majority of the currently known topological insulators are inversion symmetric, and we therefore focus on this case in the current manuscript.

\begin{figure}
	\centering
		\includegraphics[width=3.0in]{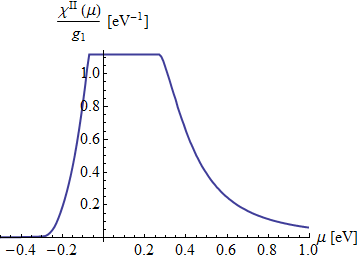}
	\caption{Plot of $\chi^{II}/g_1$ numerically calculated for \BiSe. The $\chi^{II}$ value is 
	quantized as long as the chemical potential $\mu$ is in the bulk gap. 
	Once $\mu$ is outside the gap, $\chi^{II}$ is no longer quantized.}
	\label{fig:chi_plot_1}
\end{figure}

We now turn to a concrete example for realizing $\phi$
as a spatially-varying Zeeman field in \BiSe.
The effective low energy continuum model derived for \BiSe in Ref.~\onlinecite{SCZhang:prb2010}, involves electrons in two orbitals.
originating in the p-orbitals of different atoms (Bi and Se respectively).
As a consequence, the two orbitals in general will have a different gyromagnetic ratio when coupled to a Zeeman field. Indeed, if the magnetic field $b$ is applied in the direction of
the trigonal axis of the \BiSe crystal (z-direction in the notation of Ref.~\onlinecite{SCZhang:prb2010}), one finds
$H_{\text{Zeeman}} = b \sigma_3 (g_0 + g_3 \tau_3) $,
where $\sigma_{1,2,3}$ are the Pauli matrices of the electron spin, 
and $\tau_{1,2,3}$ are the Pauli matrices describing the orbital degree of freedom. 
If the magnetic field $b$ varies on the length scale of atomic distances, 
it will effectively break inversion symmetry in the crystal and allow a more general Zeeman coupling to occur
\be\label{Zeeman}
H_{\text{Zeeman}} = b_{FM} \sigma_3 (g_0 + g_3 \tau_3) 
+ b_{AFM} \sigma_3 (g_1 \tau_1 + g_2 \tau_2) 
\; .
\ee
Here $b_{FM}$ and $b_{AFM}$ are ``ferromagnetic'' (FM) and 
``antiferromagnetic'' (AFM) fields, respectively.
The field $b_{AFM}$ is odd under \emph{both} time-reversal and inversion,
and is therefore a suitable realization of $\phi$. Generating it may require
anti-ferromagnetic order, though any magnetic order that varies strongly 
on microscopic scales (ferrimagnetism, spin spirals etc.) will suffice. 
We also note in passing that we neglect the orbital coupling of the magnetic field we apply here. With a sufficiently weak Zeeman field the flux through a unit cell of 
the solid will be small, and we can safely neglect it.  
Using \eqref{nonlinear_Bastin} we calculate $\chi^{II}$ for \BiSe,
with all the numerical parameters we need (apart from $g_1$) taken from 
Ref.~\onlinecite{SCZhang:prb2010}.
Varying the chemical potential, we find the values
plotted in Fig.~\ref{fig:chi_plot_1}.
The most striking feature in plotting $\chi^{II}$ versus chemical potential is the
\emph{plateau} in its value while $\mu$ is in the gap.
Once the chemical potential is outside the gap,
the value of $\chi^{II}$ changes continuously.
This will be the most easily discernible experimental signature of the topological insulator - a plateau in the transport coefficient $\chi$ - even though its value
is not universal.

In conclusion, we have found that the magneto-electric coupling in
topological insulators and their gapless counterparts, can be related
to a third derivative of a thermodynamic potential. Most
interestingly, this implies that the topological effects could be
measured by probing either charge-density or magnetization in
equilibrium, and in the bulk, rather than from non-equilibrium transport properties of
the surface.
Our result \eqref{grand_result}, suggests a conceptually simple 
way to measure the magneto-electric response, by measuring magnetization, 
while varying the chemical potential and the auxiliary field $\phi$, or measuring
the charge density while varying the magnetic field and $\phi$.
Moreover, our formula holds regardless of whether the system is gapless or gapped, 
clean or disordered. The measurement we propose, however, is challenging.
First and foremost, realizing the auxiliary field is difficult,
in the case of \BiSe, requiring the introduction of microscopic AFM order
to the bulk of the material, and carefully controlling it.
Controlling the chemical potential may also be challenging, given that we wish to probe 3D systems. Varying the chemical potential is needed not only to calculate the derivative in \eqref{grand_result}, but also to detect the most clear cut evidence for a topological state - the plateau 
in $\chi^{II}$, as illustrated in Fig.~\ref{fig:chi_plot_1}.
Finally, the magnetization in our formula is the \emph{orbital} magnetization, 
ignoring the Zeeman contributions to the magnetization. 
In a metal, the Zeeman contribution can vary with the chemical potential,
but in an insulator, it will not. Therefore measuring the full magnetization,
instead of the orbital magnetization, will yield $\chi^{II}$ in the insulating state, 
but in the metal it will yield $\chi^{II}$ plus some corrections.
However, the key qualitative feature is the plateau in $\chi^{II}$ 
in the insulating state, which will still show up when measuring the total magnetization instead of the orbital magnetization. It is the quantization of a response coefficient that signifies a topological incompressible state.
Despite these difficulties, our findings allow a unique conceptual 
approach to measuring the magneto-electric coupling, and it is our sincere hope 
this insight will be put to use in the lab.

\emph{Acknowledgments:} DLB was supported by the
Sherman Fairchild foundation. GR acknowledges support from DARPA and
from the Caltech Institute of Quantum Information and Matter, an NSF Physics Frontiers Center with the support of
the Gordon and Betty Moore Foundation.


\bibliographystyle{apsrev}

\end{document}